# Atomic-Scale Surface Imaging of bulk Epitaxial CsPbBr$_3$ Perovskite Single Crystals on Mica using Light Assisted Scanning Tunneling Microscopy at Low-Temperature (80 K)

Eric Duverger[1], Vladimir Bruevich[2], Vitaly Podzorov[2], Damien Riedel[3*]

[1]*Institut FEMTO-ST, Univ. Marie et Louis Pasteur, CNRS, 15B avenue des Montboucons, F-25030 Besançon, France.*

[2] *Department of Physics and Astronomy, Rutgers, the State University of New Jersey, Piscataway, New Jersey 08854, USA.*

[3]*Institut des Sciences Moléculaires d'Orsay (ISMO), CNRS, Univ. Paris Sud, Université Paris-Saclay, F-91405 Orsay, France.*

**Abstract**

Epitaxial single-crystalline CsPbBr$_3$ perovskite films on mica, prepared ex-situ, are explored using a low-temperature scanning tunneling microscope (STM) by probing the unoccupied electronic states of their surface in ultra-high vacuum (UHV) at 80 K. Light-assisted STM measurements under a broadband illumination with visible light were employed to enhance and stabilize surface conductivity. STM imaging across the surface of macroscopic bulk CsPbBr$_3$ films reveals large flat terraces characterized by a specific type of surface reconstruction, consisting of parallel rows of U-shaped atomic nanostructures. These structures are spaced by 12 Å and exhibit an internal periodicity of 5.1 Å. Density functional theory (DFT) calculations reproduce the experimental observations and reveal a competition between different orthorhombic CsPbBr$_3$(110) surface reconstructions: a Cs-rich structure, identified as the most energetically stable, and three alternative Pb–Br-rich reconstructions, which are slightly higher in energy yet remain consistent with the STM data. Additional analyses that explicitly account for the mica substrate exclude the cubic CsPbBr$_3$ phase and other orthorhombic surface orientations, while showing that variations in the mica surface termination do not alter the preferred CsPbBr$_3$(110) reconstruction. This combined approach thereby confirms our assignment and resolves previous STM interpretations of CsPbBr$_3$.

## 1- INTRODUCTION

Metal-halide perovskites (MHPs) have emerged as a transformative class of materials across a wide spectrum of scientific and technological domains[1]. These materials are characterized by an ABX$_3$ crystal structure, where an inorganic sublattice of corner sharing metal-halide octahedra (e.g., PbI$_6^-$ or PbBr$_6^-$) and another sublattice of inorganic or organic cations (e.g., Cs$^+$ or CH$_3$NH$_3^+$) together form a perovskite structure that exhibits diverse optical, electrical, and structural properties[2,3,4]. These properties have propelled applications of perovskites in photovoltaics[5,6], light-emitting diodes (LEDs)[7], photocatalysis[8], and sensors[9]. Particularly, their high absorption coefficients[10], long carrier diffusion lengths[2,11], and facile synthesis make them a versatile playground for research on the next-generation optoelectronic devices[12].

In recent years, the research focus on functional semiconducting materials and devices has significantly shifted towards the role of surfaces and interfaces because of their influence on device performance, especially at the nanoscale[13,14,15,16]. Surface states can govern charge recombination[17], stability, and interfacial interactions. Engineering surface chemistry has thus become pivotal for achieving higher efficiencies and stability in practical applications[18]. Among the broad family of halide perovskites, cesium lead bromide (CsPbBr$_3$) has drawn particular attention due to its superior thermal and structural stability as compared to hybrid (organic–inorganic) counterparts[19,20].

As a fully inorganic perovskite, CsPbBr$_3$ combines a relatively wide bandgap (~ 2.3 eV) with a high photoluminescence quantum yield and robust environmental tolerance. These characteristics make it especially suitable for blue-green light emission and stable photodetector platforms[21]. Moreover, the nanostructuring of CsPbBr$_3$ into quantum dots, nanowires, or thin films has opened exciting avenues in nanoelectronics[22], where quantum confinement effects, surface-to-volume ratios, and doping are critical[23,24].

The surface properties of CsPbBr$_3$ films play a crucial role in governing the charge transport properties, interfacial energy alignment, and defect passivation, the key aspects of high-performance electronic devices. By tailoring surface termination or introducing passivating agents, one can, in principle, improve the charge carrier mobility[25,26,27] and device stability[17]. Thus, from the perspective of the fundamental surface science and cutting-edge electronic design, CsPbBr$_3$ offers a versatile platform for basic and applied research in novel optoelectronic systems[28,29,30].

Low-temperature (LT) studies of bulk CsPbBr$_3$ are expected to benefit from reduced thermal and electronic noise, as well as improved charge transport, as cooling suppresses ionic motion, reduces phonon scattering, sharpens optical features,[31] leads to an enhanced charge carrier mobility[28], thus helping to stabilize scanning probe measurements. At cryogenic temperatures, the orthorhombic crystal structure becomes well-defined, enabling the reliable identification of surface terminations and reconstructions. Indeed, cooling CsPbBr$_3$ to 80 K stabilizes the γ-orthorhombic phase, thermodynamically favoring specific distorted PbBr$_6$ octahedral structures[32]. In this context, LT-STM can provide an enhanced spatial resolution and allows a direct mapping of surface and coordination environments with minimal thermal noise. Therefore, the insights on the nanoscale surface structure obtained with LT-STM imaging would be crucial for the understanding of the *intrinsic* (i.e., not dominated by static disorder) surface reconstruction and charge transport properties of bulk CsPbBr$_3$ including low-threshold, single-crystal CsPbBr$_3$ field-effect transistors (FETs)[28].

In this article, we investigate the atomic-scale surface structure of bulk $CsPbBr_3$ perovskite at low temperature (80 K). We use single-crystalline perovskite samples epitaxially grown on mica substrates, mounted on special sample holders compatible with LT-STM measurements. Due to the low conductivity of pristine $CsPbBr_3$ in the dark, we generate a population of charge carriers in the conduction band by illuminating the samples with a broad-band white light during STM scanning.

Imaging the bulk $CsPbBr_3$ surface with LT-STM reveals a clear and dominant U-shaped rows surface reconstruction, which is observed at the surface of macroscopically large samples. The surface exhibits step edges and coalesced step structures with relative height differences of approximately 4.8 Å. LT-STM calibration of the $CsPbBr_3$ surface reconstruction indicates a periodicity of 11.1 Å along one direction and 11.8 Å along the orthogonal direction. Through theoretical surface analysis based on density functional theory (DFT), we identify several possible variants of the $CsPbBr_3$ (110n) surface reconstructions within the orthorhombic crystal structure. These variants mainly differ in the relative surface concentrations of Cs, Br, and Pb. STM imaging of the boundaries between adjacent U-shaped domains further reveals a transitional region that connects neighboring terraces, indicating that the observed reconstructions need to accommodate domain matching at their interfaces. When the mica substrate is explicitly taken into account, both the cubic $CsPbBr_3$ phase and alternative orthorhombic surface orientations are found to be inconsistent with the STM observations. Variations in the mica surface termination do not alter the selected $CsPbBr_3$ surface reconstruction.

## 2- RESULTS AND DISCUSSION

Single-crystal $CsPbBr_3$ films were epitaxially grown on mica substrates using a custom-built growth system (for details, see Methods and Ref. 28). The samples were handled and stored in ambient air; for long-term storage/shipping, the samples were kept in individual light-proof containers (Fig. 1a). Before STM measurements, the samples were pre-cut to rectangular pieces of desired dimensions, and graphitic contacts were painted at two opposite ends of each sample (see black rectangles in Fig. 1a). This step was necessary to mount the $CsPbBr_3$ samples onto a dedicated sample holder compatible with our low-temperature STM system (Fig. 1b) and further ground the samples. Since the $CsPbBr_3$ samples were exposed to ambient air prior to their introduction into the UHV preparation chamber, we gently degas them before performing STM measurements. To achieve this, we first place a rectangular piece of a sacrificial doped Si(100) wafer onto the two molybdenum blocks of the sample holder (Fig. 1c), and then place the pre-cut $CsPbBr_3$ sample on top of it (Fig. 1d). The assembly is then clamped together using molybdenum hooks screwed into place on both sides of the sample holder (Fig. 1e) ensuring a good electrical connection between the graphitic contacts and the Si substrate. In this configuration, Si provides a flat, relatively low-resistance (~ 15 Ω), grounded substrate that supports the much more resistive (0.1 - 1 MΩ) $CsPbBr_3$ film, with both graphitic contacts grounded, thus bypassing the insulating mica substrate.

One can note the yellowish appearance of the $CsPbBr_3$ films (Fig. 1d,e). Bulk $CsPbBr_3$ can adopt two main crystallographic structures. First is an orthorhombic phase with a unit cell measuring 12.07 Å × 8.4 Å × 8.4 Å (Fig. 1f). For bulk free-standing $CsPbBr_3$ crystals, this structure is stable between very low temperatures (~ 4 K) and up to approximately 361 K[33]. The second structure is a face-centered cubic (fcc) phase with a unit cell size of 5.6 Å (Fig. 1g), which is generally stable above ~ 405 K. These unit cell dimensions will serve as a reference point for comparing the surface reconstruction observed in our $CsPbBr_3$ films by STM measurements at LT. Note that an intermediate tetragonal phase may also exist between these two main phases[34,24].

Our preliminary investigation using polarized optical transmission microscopy reveals that the samples consist of two main areas of different contrast suggesting that the surface reconstruction may be of two distinct kinds (Fig. S1a). Due to the stability of the γ-orthorhombic phase at LT, the cubic phase can usually be ruled out as a structural origin for $CsPbBr_3$ surface terminations at LT. Considering the various growth orientations on mica reported in the literature, we envisioned the (100), (010) and (110) surface terminations[35]. However, our STM investigation performed at multiple spots revealed only a single surface reconstruction.

Prior to STM imaging, we measured the band gap of the $CsPbBr_3$ samples to determine a suitable bias voltage for optimal imaging conditions. This was done using photoluminescence (PL) spectroscopy (Figs. S1b, S1c). Under ambient laboratory light, the $CsPbBr_3$ sample shows no significant PL emission (Fig. S1b). In contrast, a bright green emission is clearly observed when the film is excited with a 405 nm laser diode (Fig. S1d, shown in air). A spectrometer was then used to collect and analyze the emitted green PL once the sample is in the STM chamber (Fig. 2a). The resulting spectrum, shown in Fig. S1d, reveals that the PL of bulk $CsPbBr_3$ on mica is centered at 532.6 nm with a relatively narrow bandwidth of approximately 17 nm (at FWHM), corresponding to an apparent band gap of $2.33 \pm 0.04$ eV[36].

To investigate surface reconstruction at the atomic scale, good surface conductivity under our experimental conditions is essential. High-purity (undoped) bulk $CsPbBr_3$ on mica samples exhibit a room-temperature resistivity in the dark in the range of $0.4 - 0.7$ GΩ·cm[37], corresponding to a low conductivity of approximately $1.9 \times 10^{-9}$ S/cm. For comparison, the sacrificial silicon substrates used in these experiments have a much lower resistivity of $5 \times 10^{-3}$ Ω·cm[38,39]. Although the intrinsic carrier mobility of $CsPbBr_3$ increases upon cooling[28], the concentration of mobile carriers available in the conduction band of pristine crystals is very low in the dark, particularly for bulk $CsPbBr_3$ supported on mica. Consequently, light-assisted STM imaging significantly enhances intrinsic carrier transport through the generation of additional carriers in the conduction band, further aided by the γ-orthorhombic phase, which stabilizes defect dynamics and suppresses lattice disorder[32]. An improved mobility at LT is important as it increases the surface conductivity of the sample in the situation of low carrier concentration, ensuring a sufficient carrier diffusion away from the tip-sample junction without trapping. Although reliable data on the low-temperature conductivity of bulk $CsPbBr_3$ are scarce due to sample-to-sample variations in purity, the conductivity of high-purity bulk $CsPbBr_3$ at low temperature is expected to be very low. Our measurements show that STM imaging of pristine bulk $CsPbBr_3$ surface in the dark is not feasible at low temperature, in contrast to studies on thin $CsPbBr_3$ films grown on metal surfaces[40], unless additional free carriers are photo-generated.

Following a mild degassing of the $CsPbBr_3$ samples for ~ 2 hours at 60°C (see Methods), the sample holder was introduced in the STM chamber pre-cooled via a dilution cryostat (Fig. 2a). The ultimate pressure reached in the STM chamber is reduced due to a cryogenic pumping effect, allowing for efficient and stable surface imaging to be performed. STM imaging was tested with and without photoexcitation; in the dark, tip stabilization was achievable, but scanning induced severe feedback instability, preventing atomic-resolution imaging. Under these conditions (80 K, in the dark), the concentration of mobile carriers at the surface of $CsPbBr_3$ was insufficient to ensure good electrical connectivity between the two electrodes of the sample, therefore, light-assisted techniques were required[41,42]. Notably, varying the STM tip's bias polarity (positive or negative) did not lead to any improvement. LT-STM imaging was attempted using the same 405 nm laser diode employed for PL measurements (Fig. S1d). However, the tunneling current exhibited large fluctuations during scanning, which degraded the signal-to-noise ratio and prevented atomic-scale resolution.

To take full advantage of the high intrinsic electronic quality of our CsPbBr$_3$ samples and the favorable carrier mobility of the γ-orthorhombic phase, we used illumination from a broadband white lamp, whose emission spectrum is shown in Fig. 2b. This illumination penetrates more deeply into the sample and reduces tunneling-current fluctuations. Under these conditions, light-assisted STM imaging at LT benefits from enhanced surface conductivity, as photo carrier generation and reduced non-radiative recombination are observed to stabilize the tunneling current as shown in the I(V) curve in Fig. 2c. Indeed at LT, the γ-orthorhombic phase shows a higher photoluminescence quantum efficiency due to suppressed phonon modes, which reduce defect-assisted recombination and extend the carrier lifetime[43]. Under this illumination scheme, we were able to successfully image the bulk CsPbBr$_3$ surface on mica at LT. Under our experimental conditions, highly localized dI/dV spectroscopy does not provide meaningful information on the elemental composition of the CsPbBr surface, owing to its large surface bandgap.

Once the CsPbBr$_3$ sample is introduced into the STM, we have acquired the first image of a large surface area (Fig. 3a). Residual surface contamination can hinder stable atomic-scale imaging of the surface reconstruction over large areas, although step-edges between adjacent terraces remain clearly resolved. A profile taken across the STM image (blue rectangle in Fig. 3a) allows us to estimate the height ~ 4.85 Å between the two observed terraces and a relatively low corrugation (~ 1 Å) at the terraces (Fig. 3c). A magnified view of the same region (red square in Fig. 3a) reveals an STM image displaying a wide area with atomic-scale resolution of the surface reconstruction (Fig. 3b). The surface appears structured with parallel rows composed of periodic chevron patterns. Although residual contamination limits the extent of fully clean regions displaying this reconstruction, the observed surface structure contains very few surface defects. Additional scans performed at different surface locations reveal coalescing step edges and parallel terraces, both exhibiting similar step heights of ~4.81 Å and 4.82 Å (Figs. 3d and 3e).

We explored various regions of the $CsPbBr_3$ surface across the macroscopic sample and consistently detected a single dominant surface reconstruction. Similar to the one firstly identified in Fig. 3b, this reconstruction is examined in detail in Figs. 4a and 4b. Figure 4a shows a terrace of the $CsPbBr_3$ surface observed during our measurements. It reveals a distinct U-shaped pattern, periodically spaced along one direction. In the perpendicular direction, these U-shaped rows run parallel to one another. The edge of the imaged terrace appears disordered, a consequence of the surface-atom diffusion expected during epitaxial growth. Fig. 4b shows a zoomed-in region from Fig. 4a (blue rectangle), detailing the surface structure over an area of approximately $5 \times 4$ unit cells. We have acquired several profiles on this area along the $\overrightarrow{XX'}$ and $\overrightarrow{YY'}$ directions. The averaged profiles are shown in Fig. 4c. Perpendicular to the rows, the profile along the $\overrightarrow{XX'}$ direction exhibits a main periodicity of 11.1 Å and an apparent second periodicity inside the U-shaped structure of 5.1 Å. The observed periodicities along $\overrightarrow{YY'}$, which corresponds to the orientation of the U-shape rows, have a slightly greater main periodicity at 11.8 Å and the secondary periodicity remaining at 5.1 Å.

This surface structure, composed of periodically arranged U-shaped rows, also appears in specific cases where two terraces of identical height merge (Fig. 4d). In Fig. 4d, regions $z_1$ and $z_3$ both display the same periodic U-shaped row pattern, separated by a narrow interfacial area $z_2$, which exhibits a slightly different structure with short staggered dashes. Interestingly, the orientation of the U-shaped rows in $z_1$ and $z_3$ differs by an angle of approximately 108° (Fig. 4e). Previous studies of thin $CsPbBr_3$ films surface reconstructions on metal substrates report patterns similar to the one observed in the interface region $z_2$[44,45,46]. Often called armchair structures, they appear clearly different from the U-shaped observed on bulk $CsPbBr_3$ samples on mica under STM imaging. The same trend is also observed for orthorhombic MAPbBr$_3$(010) surface or when Pb atoms were substituted with Sn[47,44,45]. A more detailed analysis of these surface reconstructions on gold is discussed in Fig. S2 and note N1.

To better understand the origins of the observed surface patterns, we have performed a theoretical analysis based on DFT to characterize possible reconstructions of the CsPbBr$_3$ surface, considering the orthorhombic unit cell shown in Fig. 1f, cleaved along the (110) plane. Additional tests based on the cubic and orthorhombic CsPbBr$_3$ unit cells for various surface orientations ((001), (010), and (110), Fig. S3 and S4), show no satisfactory agreement with the experimental data and can therefore be excluded as descriptions of the observed U-shape rows on the bulk CsPbBr$_3$ surface structure on mica.

The successful DFT simulations are presented in Fig. 5, illustrating four surface structures, ranked by increasing total energy, that best match our experimental observations. The four surface structures obtained are characterized by either Cs termination or Pb-Br termination. These results align with existing literature, which identifies two distinct types of charge-neutral surfaces formed when the orthorhombic CsPbBr$_3$ crystal is terminated by its (110) facets[34].

Each case, in Fig. 5, is shown from left to right as follows: a top view of four surface unit cells, a side view of the same structure, a top view of the first three atomic layers of a single unit cell, its corresponding side view, and finally, the calculated local density of states (LDOS) for that configuration. Although each modeled unit cell follows the stoichiometry Cs$_8$Pb$_8$Br$_{24}$ (i.e., CsPbBr$_3$), a more precise description of each surface reconstruction is provided by considering the first three atomic layers and labeling each case accordingly as Cs$_n$Pb$_m$Br$_p$.

The CsPbBr$_3$ surface structure shown in Fig. 5a appears to be the most stable configuration ($E_{tot.}$ = –25955.52 eV) and corresponds to a Cs-rich phase, characterized by two equally spaced Cs adatoms that primarily account for the observed surface periodicity, as reflected in the calculated local density of states (LDOS). The LDOS image closely resembles the U-shaped rows observed experimentally. Considering the first three atomic layers of the unit cell, the reduced surface composition corresponds to $Cs_2Pb_4Br_8$. The other three configurations (Figs. 5b–5d) represent Pb-Br–rich surfaces, with Cs atoms located deeper within the structure. For clarity, we show their positions superimposed on the LDOS images, although they belong to the fourth atomic layer. The surface structures in Figs. 5b and 5d both exhibit LDOS patterns that reproduce very well the U-shaped rows seen in Fig. 4. In Fig. 5b, the U-shape is more pronounced than in Fig. 5a, and this effect becomes even more prominent in Fig. 5d.

The key distinction between these three Pb-Br-rich surface reconstructions (Figs. 5b – 5d) lies in the ratio of Br to Pb atoms varying across their top three atomic layers. The reduced surface structure in Fig. 5b is $Cs_0Pb_4Br_8$, while Fig. 5c shows $Cs_0Pb_4Br_6$, and Fig. 5d has a $Cs_0Pb_4Br_4$ reduced structure. This progression indicates that achieving parity between lead and bromine contents at the very first atomic layers results in a less favorable surface energy. Interestingly, the U-shaped pattern change in the calculated structure shown in Fig. 5c. This configuration differs from Fig. 5b and Fig. 5d only in its surface composition, $Cs_0Pb_4Br_6$, which leads to a non-integer Br/Pb ratio.

A comparison between the LDOS image and the interface structure observed in region $z_2$ of Figs. 4d and 4e suggests a potential match with the structure presented in Fig. 5c. Notably, the periodicity *e* of the short staggered dashes in $z_2$ is the same as that of the U-shaped rows in regions $z_1$ and $z_3$ (Fig. 4e). This reconstruction has been reported as one of the dominant $CsPbBr_3$ surface phases on metals (armchair structure), observed for ultrathin films grown on Au(001) and Au(111). Remarkably, the same armchair pattern with nearly identical periodicities was reported over a wide bias range, at both room and low temperature, and irrespective of whether a cubic or an orthorhombic unit cell was assumed[45]. Since these two lattices are crystallographically incompatible, such invariance cannot be intrinsic to $CsPbBr_3$ and instead indicates a strong templating effect from the metallic substrate. Instead, our present study of bulk $CsPbBr_3$ films on mica shows that even slight variations in the surface Pb/Br atomic ratio can affect STM image interpretation.

Our theoretical investigations, detailed in Fig. 5, indicate that the most stable $CsPbBr_3$ surface phases at low temperatures are either Cs-rich or Pb/Br-rich. Further theoretical data, consistent with existing literature, indicate that the Cs-rich surface is energetically more stable than the Pb-Br-rich surface[48]. At first glance, this energetic preference may arise because terminating the surface at the $PbBr_2$-rich side requires breaking the $PbBr_6$ octahedron, making it energetically less favorable compared to termination at the CsBr-rich surface[34]. However, comparison between the experimental STM images and the calculated LDOS images of the (110) surface of the $CsPbBr_3$ film suggests that this surface is predominantly composed of a Pb-Br-rich surface reconstruction with reduced surface structures ranging between $Cs_0Pb_4Br_8$ and $Cs_0Pb_4Br_6$. While our data do not allow for a definitive distinction between Cs-rich and Pb-Br-rich surface phases, the overall characteristics of the U-shape surface reconstruction observed in this study strongly suggest that the Pb-Br–rich $Cs_0Pb_4Br_8$ structure corresponds to the U-shaped row reconstruction imaged by STM at LT. One can note that the calculated structure in Fig. 5b reproduces the step edge height (~4.8 Å) as well as the trend of the second periodicity observed in the STM profiles (Fig. 4c). The slight discrepancy between the measured value (5.1 Å) and the corresponding value in the calculated structure (5.8 Å) may result from a relative shift between the actual atomic positions and the density of states maximum detected by the STM tip at the considered bias.

Additional investigations of the $CsPbBr_3$(110) structure adsorbed on muscovite mica further support our results (Fig. 6). Using a monoclinic $[KAl_2(SiO_3)_4]_2$ unit cell (Fig. 6a), the (001) surface slab can be reproduced, consistent with the natural cleavage of mica along the basal

(001) plane, where the angle between the crystallographic a and b axes is 60° (Fig. 6b). In addition to the Al/O termination, two main mica (001) surface terminations were examined, including Si/O and K/O terminations, all described using a 2 × 2 unit cell (10.44 Å × 10.44 Å; Fig. S5). Several orthorhombic CsPbBr$_3$(110) surface reconstructions adsorbed on mica were investigated. Figures 6c, 6f and 6i illustrate the Cs$_2$Pb$_4$Br$_8$(110) Cs-rich termination, as defined in Fig. 5a, grown on Al/O-terminated mica (001) (Fig. 6i). This configuration allows to reproduce the calculated STM LDOS image (Fig. 6i) as observed in Fig. 5a (right panel).

Figs. 6d, 6g, 6j and 6e, 6h, 6k exhibit the Cs$_0$Pb$_4$Br$_4$(110) and Cs$_0$Pb$_4$Br$_8$(110) surfaces, with Al/O and Si/O terminated mica (001) surfaces, respectively. In both cases, the calculated STM images are consistent with the results previously reported in Figs. 5d and 5b. These findings further indicate that the mica surface termination does not significantly affect the final stoichiometry of the resulting CsPbBr$_3$(110) surface. With mica, the surface unit cell of the calculated CsPbBr$_3$(110) structure undergoes a slight distortion, increasing from 12.00 to 12.04 Å along the [001] direction and decreasing from 11.90 to 11.74 Å along the [$\bar{1}$10] direction. Further analysis of the relative orientation between the CsPbBr$_3$ overlayer and the mica substrate reveals a small angular mismatch of ~ 6° (Fig. S6). This offset supports our previous conclusions and accounts for the 108° angle observed between adjacent domains in Fig. 4e: two successive mica directions separated by 120° give rise to CsPbBr$_3$ (110) domains rotated by 108° (120° − 2 × 6°).

### 3- CONCLUSIONS

Using low-temperature light-assisted STM imaging, the surface reconstruction of epitaxially grown single-crystal CsPbBr$_3$ perovskite films was investigated. We observed that the dominant surface reconstruction, characterized by periodic rows, extends across large terraces with few defects. Each row is composed of repeating U-shaped units, with a face-to-face arrangement between adjacent rows. DFT calculations support our experimental observations and suggest four main surface structures that can reproduce the STM images: a Cs-rich reconstruction ($Cs_2Pb_4Br_8$) with the lowest formation energy, and three Pb-Br-rich reconstructions (namely, $Cs_0Pb_4Br_{8/6/4}$). Highlighting these differences emphasizes the critical role of Br adatoms in stabilizing the surface of this perovskite system.

We also demonstrated that LT-STM (80 K) imaging of pristine bulk $CsPbBr_3$ surfaces supported on mica is feasible under a broadband (white-light) photoexcitation despite the very low conductivity of the thick perovskite pristine films in the dark. LT-STM measurements confirm the presence of the orthorhombic phase and enable more precise surface characterization. This work reveals the fundamental surface reconstruction of $CsPbBr_3$ films, laying the groundwork for a deeper understanding of surface and interface phenomena in perovskite optoelectronic devices, such as interfacial charge transport, carrier trapping, and excitonic processes. Our LT-STM investigation suggests that atoms at the as-grown surface of epitaxial $CsPbBr_3$ crystals are fully coordinated, with nearly all bonds completed. This leads to the surface with a few dangling bonds, and hence a low density of deep traps - a key feature allowing a successful fabrication of intrinsic (i.e., not dominated by traps) FETs on as-grown $CsPbBr_3$ surfaces[28].

## 4. METHODS

### 4.1 Vapor-phase epitaxy of single-crystalline $CsPbBr_3$ perovskite films

Cesium bromide (CsBr) and lead(II) bromide (PbBr$_2$) precursors of 99.999% purity (Sigma-Aldrich) were mixed at a 1:1 molar ratio and reacted by heating this stoichiometric mixture in an alumina boat at 380 °C for 12 h in an inert atmosphere. Mineral muscovite mica (Electron Microscopy Sciences) was cleaved in a regular laboratory air just before the growth. The mica substrates and the alumina boat filled with a powder of the source material were loaded into a custom-designed chemical vapor deposition (CVD) growth reactor consisting of a quartz tube in a tube furnace. Ultra-high purity He was used as a carrier gas. The growth was performed at a gas flow rate and pressure stabilized at 100 sccm and 0.1 bar, respectively. The source was melted at 560 °C, while the substrate was initially heated to 500 °C, leading to a sublimation of contaminants off its surface. An epitaxial growth of CsPbBr$_3$ was initiated by shifting the substrate's position with respect to the temperature gradient to slightly reduce the substrate's temperature. The typical duration of the growth was 15 min. The growth was then abruptly terminated by quickly withdrawing the source from the hot zone of the furnace without changing the position of the substrate. Under these conditions, the tested samples have an estimated thickness ranging from 300 to 1000 nm and can be considered bulk CsPbBr$_3$ films. It has also been previously demonstrated that epitaxial growth under these conditions yields nearly stoichiometric crystalline CsPbBr$_3$ films. Subsequently, the system was gradually cooled to room temperature. More details on epitaxial growth of CsPbBr3 can be found elsewhere[28].

**4.2 Preparation of the CsPbBr$_3$ perovskite samples for STM imaging**

Once the CsPbBr$_3$ samples are mounted on the sample holder as shown in Figs. 1c–1e, they are transferred into the UHV preparation chamber for gentle degassing. Degassing is performed by heating the CsPbBr$_3$ sample, which is achieved indirectly through resistive heating of the underlying silicon substrate. A voltage bias of a few volts is applied while regulating the current through the silicon sample to control the temperature. The calibration of the silicon surface temperature is performed in a separate measurement, conducted without the $CsPbBr_3$ sample on top. Based on this calibration, the $CsPbBr_3$ samples are heated to approximately 60 °C for about 2 hours before any measurements are taken. After degassing, the sample holder is transferred into the STM chamber, which is stabilized at 80 K.

### 4.3 Theoretical tools to simulate the CsPbBr3 surface structure

For surface slab generation of $CsPbBr_3$, we use the Pymatgen software through the precise cleavage of the bulk orthorhombic crystal to obtain (100) and (110) surfaces, directly mirroring common experimental sample orientations.[35] We have therefore obtained three distinct surface terminations for (100) surface and four distinct surface terminations for (110) surface. For each of these configurations, we carefully construct surface slabs of ~14 Å depth, ensuring they contain a sufficient number of atomic layers to accurately represent the bulk while incorporating a vacuum spacing (20 Å) to minimize undesirable interactions between periodic images. Following surface generation, structural relaxation of the $CsPbBr_3$ surface slabs was a crucial next step. We have performed these relaxations using the SIESTA DFT code, which effectively utilizes numerical atomic orbitals as basis sets. Our different systems comprised 40 atoms, representing the three distinct species: Cs, Pb, and Br. We hence opted for a Double-Zeta Polarized basis set, which provides a robust description of the electronic states. We selected the Van der Waals (VDW) functional with the DRSLL parametrization for the exchange-correlation functional. Furthermore, calculations were performed with spin polarization that enables to account for any potential magnetic effects within the system. For precision, we used a real-space mesh cutoff of 150 Ry, which precisely defined the fineness of the real-space integration grid, guaranteeing high accuracy in our potential and charge density calculations. To accurately sample the Brillouin zone and ensure reliable electronic properties for our periodic surface system, we employed a 3x3x3 Monkhorst-Pack k-point grid giving 18 k-points. Atomic positions were relaxed using the Conjugate Gradient (CG) algorithm. Convergence was deemed complete once the maximum force on any atom fell below 0.02 eV/Å and the electronic self-consistent field (SCF) reached an energy convergence of $3\times10^{-2}$ eV, ensuring a truly stable configuration. Finally, thanks to relaxed surface structures, we generate theoretical scanning tunneling microscopy (STM) images by calculating the local density of states (LDOS) within a specific energy window, ranging from the Fermi level energy $E_f$ up to $E_f + 4$ eV. To note, the chosen energy range corresponds to the tunneling of electrons into unoccupied states, thereby accurately simulating the operation of a scanning tunneling microscope (STM) under positive bias. These simulated images serve as a direct benchmark for comparison with experimental observations, allowing for the (100) and (010) orientations to be disfavored.

We also investigated the interfacial properties of bulk $CsPbBr_3$ films deposited on muscovite mica. This heterostructure is modeled by a large-scale interface slab comprising 1240 atoms, specifically aligning the orthorhombic $CsPbBr_3$(110) crystallographic orientation with the (001) basal plane of the muscovite mica. The resulting supercell dimensions were defined as $26.02 \times 27.16 \times 35.89$ Å, providing a sufficiently large lateral area to accommodate the epitaxial relationship and minimize strain artifacts. To maintain consistency with our previous surface calculations, these large-scale simulations were performed using the same simulation constraints. We used the Van der Waals (VDW) functional with DRSLL parametrization and a Double-Zeta Polarized (DZP) basis set to accurately capture the non-covalent interactions at the perovskite-mica interface. The electronic environment was sampled using a k-grid configuration resulting in 18 k-points, ensuring a converged description of the density of states for the 1240-atom system. All other numerical parameters, including the 150 Ry real-space mesh cutoff, spin polarization, and the 0.02 eV/Å force convergence criteria, were kept identical to the surface slab studies. This rigorous approach allows for a direct comparison between the electronic properties of the bulk surfaces and the substrate-supported thin films, providing a comprehensive understanding of the structural stability and the resulting STM signatures of $CsPbBr_3$ on muscovite mica.

The relative stability of the considered $CsPbBr_3$(110) surface reconstructions was assessed using DFT calculations performed on slab models with identical size and stoichiometry. Total energies for each slab were compared in conjunction with simulated local densities of states (LDOS), allowing a direct and physically consistent comparison with STM measurements. This combined energetic and electronic-structure approach provides a robust criterion for identifying the most relevant surface reconstruction.

## 5- ASSOCIATED CONTENT
### 5.1 Authors' information

Dr Eric Duverger, ORCID N° 0000-0002-7777-8561, eric.duverger@univ-fcomte.fr

Dr. Damien RIEDEL, ORCID N° 0000-0002-7656-5409, damien.riedel@universite-paris-saclay.fr

Vladimir Bruevich, ORCID N° 0000-0003-3643-5391, bruevich@physics.rutgers.edu

Vitaly Podzorov, ORCID N° 0000-0001-8276-882X, podzorov@physics.rutgers.edu

**\*** Corresponding author: Dr. Damien RIEDEL, damien.riedel@universite-paris-saclay.fr, web site: http://mnd-sciences.com

### 5.2 Acknowledgements

DR would like to thank CNRS for the grant of an International Joint Research Projects (PICS) program. ED would like to thank the Mesocentre de calcul of the Franche-comté University and the *Communauté d'Agglomération du Pays de Montbeliard* (convention PMA-UFC). VP and VB thank the financial support of their part of work at Rutgers University by the U.S. Department of Energy, Office of Basic Energy Sciences, Division of Materials Sciences and Engineering under Award DE-SC0025401 (material synthesis).

### 5.3 Author contributions

DR carried out the STM experiments, collected and analyzed the data, created the figures and mainly wrote the article. ED performed the DFT simulations, including slab relaxations and LDOS images. VB and VP synthesized perovskite samples. All the authors contributed to editing and proofreading the manuscript.

### 5.4 Competing interests

The authors declare no competing interests.

**Figure captions:**

**Figure 1:** (a) Epitaxial single-crystal $CsPbBr_3$/mica samples in a storage/shipping container. Graphite contacts (black regions) on the perovskite surface (yellow regions) can be seen. (b) STM sample holder with a $CsPbBr_3$ sample mounted on it. (c - e) Details of the three-step mounting procedure of $CsPbBr_3$ samples on the STM sample holder: (c) installing an n-doped Si(100) sample on molybdenum blocks; (d) installing a $CsPbBr_3$/mica sample on top of the Si(100) surface; (e) fixing the two samples (Si(100) and $CsPbBr_3$/mica) with molybdenum clamps. (f) and (g) Ball-and-stick models of the orthorhombic and cubic unit cells of $CsPbBr_3$, where the green, brown, and dark gray balls represent Cs, Br, and Pb atoms, respectively.

**Figure 2:** (a) simplified sketch of the experimental setup. (b) Emission spectrum of a broadband, white-light lamp (Edmund LS-WL1, 1W, 0.1 mm²) used for light assisted LT STM imaging of CsPbBr$_3$ surface at 80 K. (c) I(V) curves acquired on the bulk orthorhombic CsPbBr$_3$(110) on mica, with (black curve) or without (red curve) light.

**Figure 3:** (a) example of a 873 x 873 Å² STM topography ($V_s$ = 4.0 V, I = 15 pA) acquired on the CsPbBr$_3$ surface showing two terraces of different height. (b) 377 x 377 Å² STM topography ($V_s$ = 3.8 V, I = 15 pA) acquired on the CsPbBr$_3$ surface as a detail of the red square shown in (a). The U-shaped rows can be clearly observed on both terraces. (c) Relative height profile acquired across the surface in (a) (blue rectangle). (d) 490 x 490 Å² STM topography ($V_s$ = 4.0 V, I = 19 pA) of the CsPbBr$_3$ surface showing two surface step edges, with one of them arising from a step-edge coalescence. (e) 3D representation of the two-step edges shown in (d). The relative height differences between the step edges is ~ 4.8 Å.

**Figure 4:** (a) 145.6 x 130.6 Å² STM topography ($V_s$ = 4.0 V, I = 15 pA) acquired on the CsPbBr$_3$ surface showing the details of the surface reconstruction. (b) 57.8 x 49.6 Å² STM topography ($V_s$ = 4.0 V, I = 15 pA) acquired on a small area in (a) (blue rectangle). (c) Relative height variations along the profile curves in the XX' and YY' directions, respectively. (d) and (e) 293.2 x 68.3 Å² STM topography ($V_s$ = 4.0 V, I = 15 pA) acquired across the CsPbBr$_3$ surface showing the intermediate zone ($z_2$) that connects together two other zones ($z_1$ and $z_3$) exhibiting surface reconstruction with the U-shaped periodic rows. In (e), the lattice period *e* of each structure is recalled. The relative angle between the orientations of zones $z_1$ and $z_3$ is ~ 108°.

**Figure 5:** (a) to (d) Calculated structures of CsPbBr$_3$ surface reconstruction in 4 different cases of increasing energy. From left to right: ball-and-stick top view of a four unit cells surface structure, ball-and-stick side view of the same structure, ball-and-stick top view of the first three layers of a single unit cell (the size of the atoms is divided by two for clarification), ball-and-stick side view of the reduced unit cell, gray scale calculated local density of states (LDOS) image of the corresponding structure of E$_f$ at E$_f$ + 4 eV. The reduced top-view unit cell is superimposed on the LDOS image for comparison. The green, brown, and dark gray balls represent Cs, Br, and Pb atoms, respectively.

**Figure 6:** (a) Ball and stick representation of a mica monoclinic [KAl$_2$(SiO$_3$)$_4$]$_2$ unit cell. (b) Al/O terminated mica slab (001) surface made of 5 x 5 unit cells as presented in (a) used for DFT simulations. (c),(f) and (i) ball-and-stick top view, side view and space-filling top view, respectively, of CsPbBr$_3$ terminated as Cs$_2$Pb$_4$Br$_8$(110) surface adsorbed on Al/O terminated mica. (d),(g) and (j) ball-and-stick top view, side view and space-filling top view, respectively, of CsPbBr$_3$ terminated as Cs$_0$Pb$_4$Br$_4$(110) surface adsorbed on Al/O terminated mica. (e),(h) and (k) ball-and-stick top view, side view and space-filling top view, respectively, of CsPbBr$_3$ terminated as Cs$_0$Pb$_4$Br$_8$(110) surface adsorbed on Al/O terminated mica. In (i), (j) and (k) are superimposed the calculated LDOS image with a transparency of 15%.

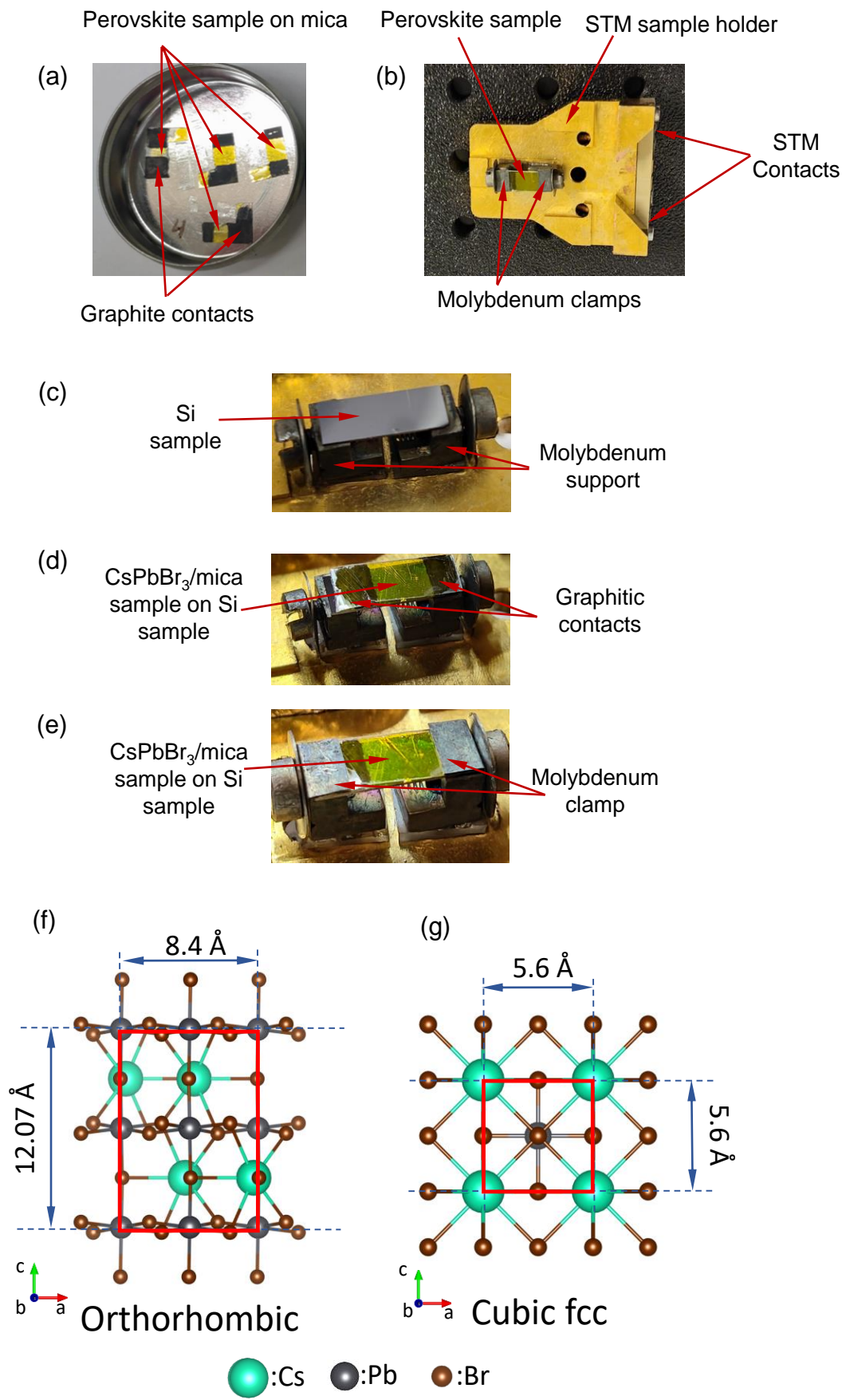

Riedel et al. Figure 1

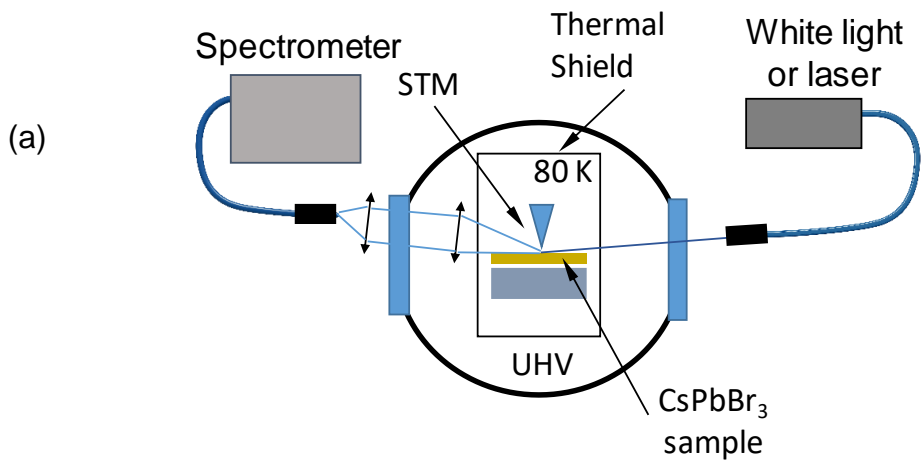

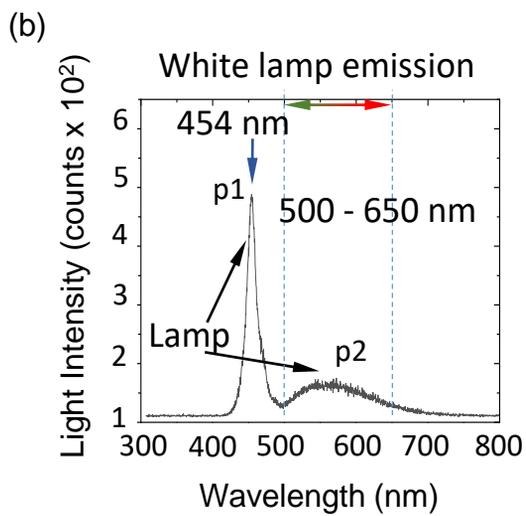
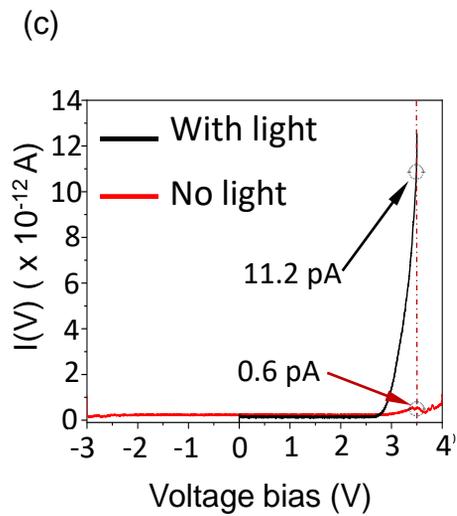

Riedel et al. Figure 2

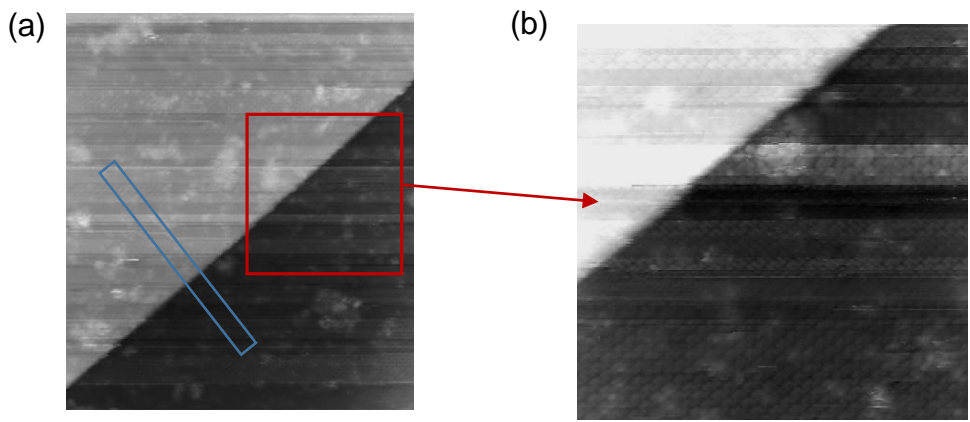

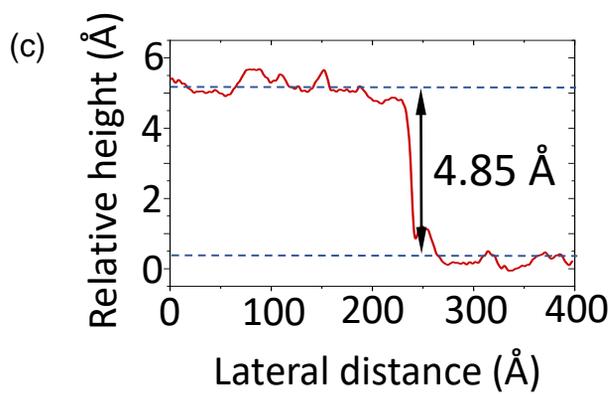

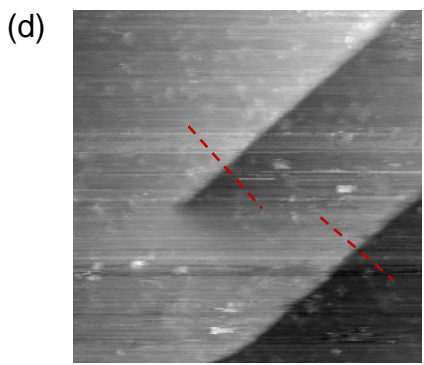 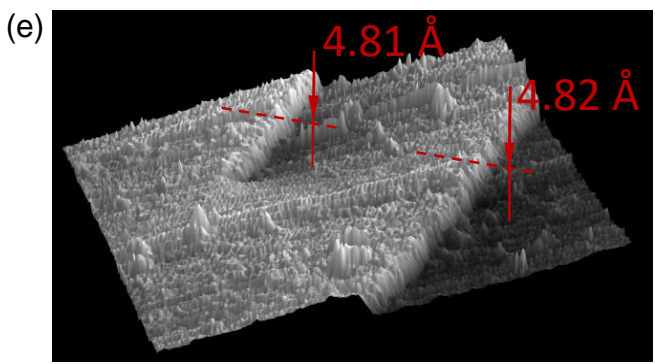

Riedel et al. Figure 3

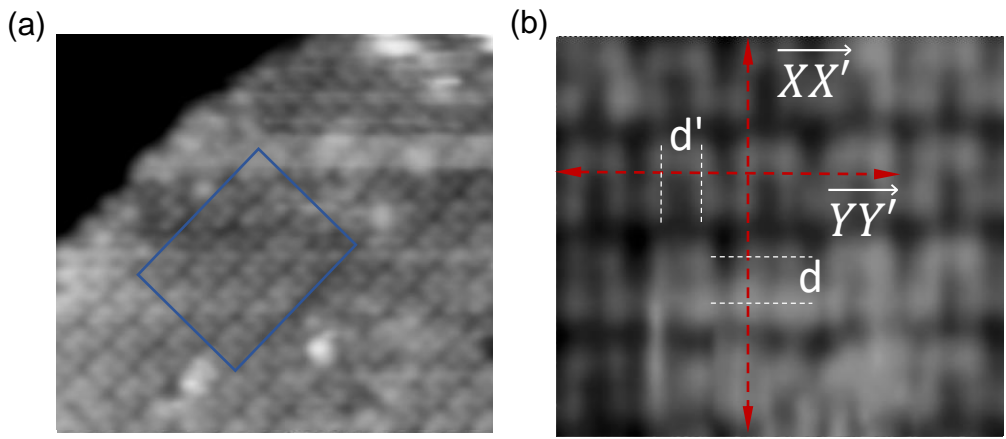

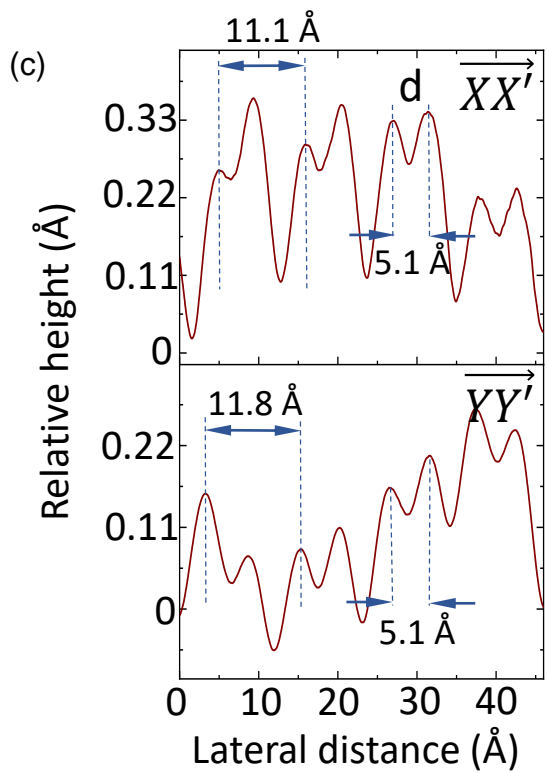

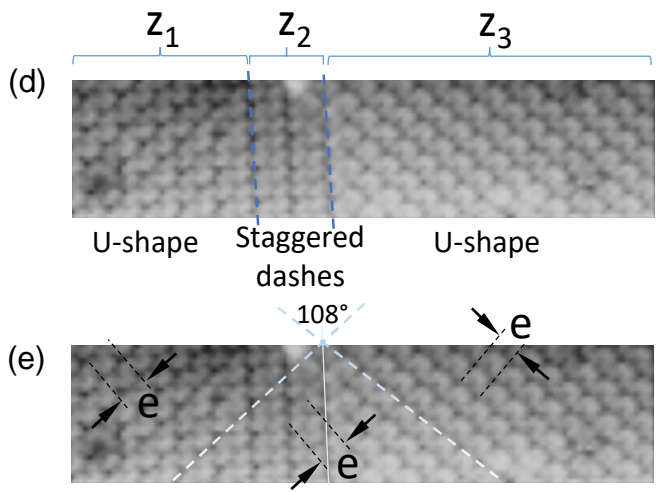

Riedel et al. Figure 4

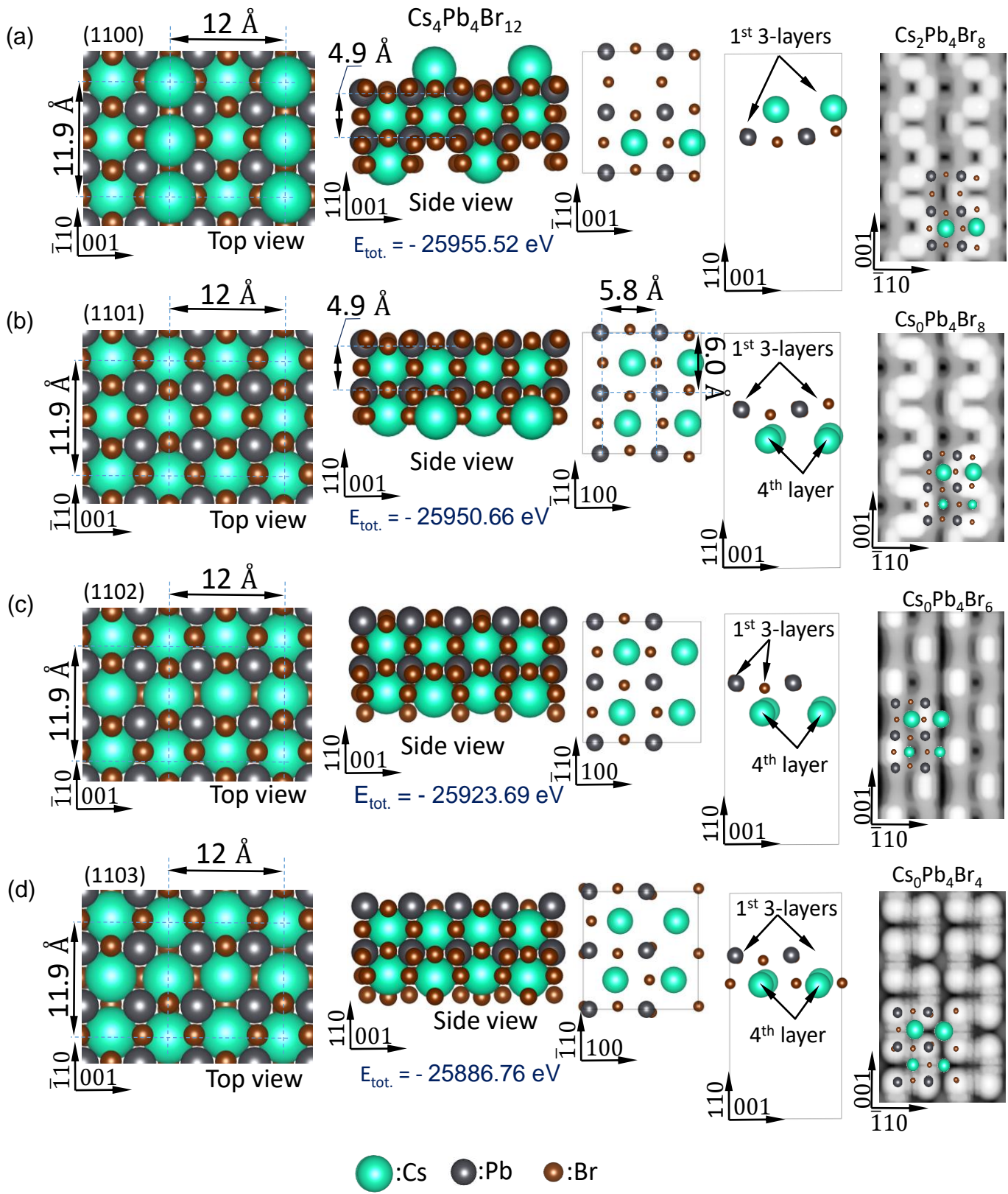

Riedel et al. Figure 5

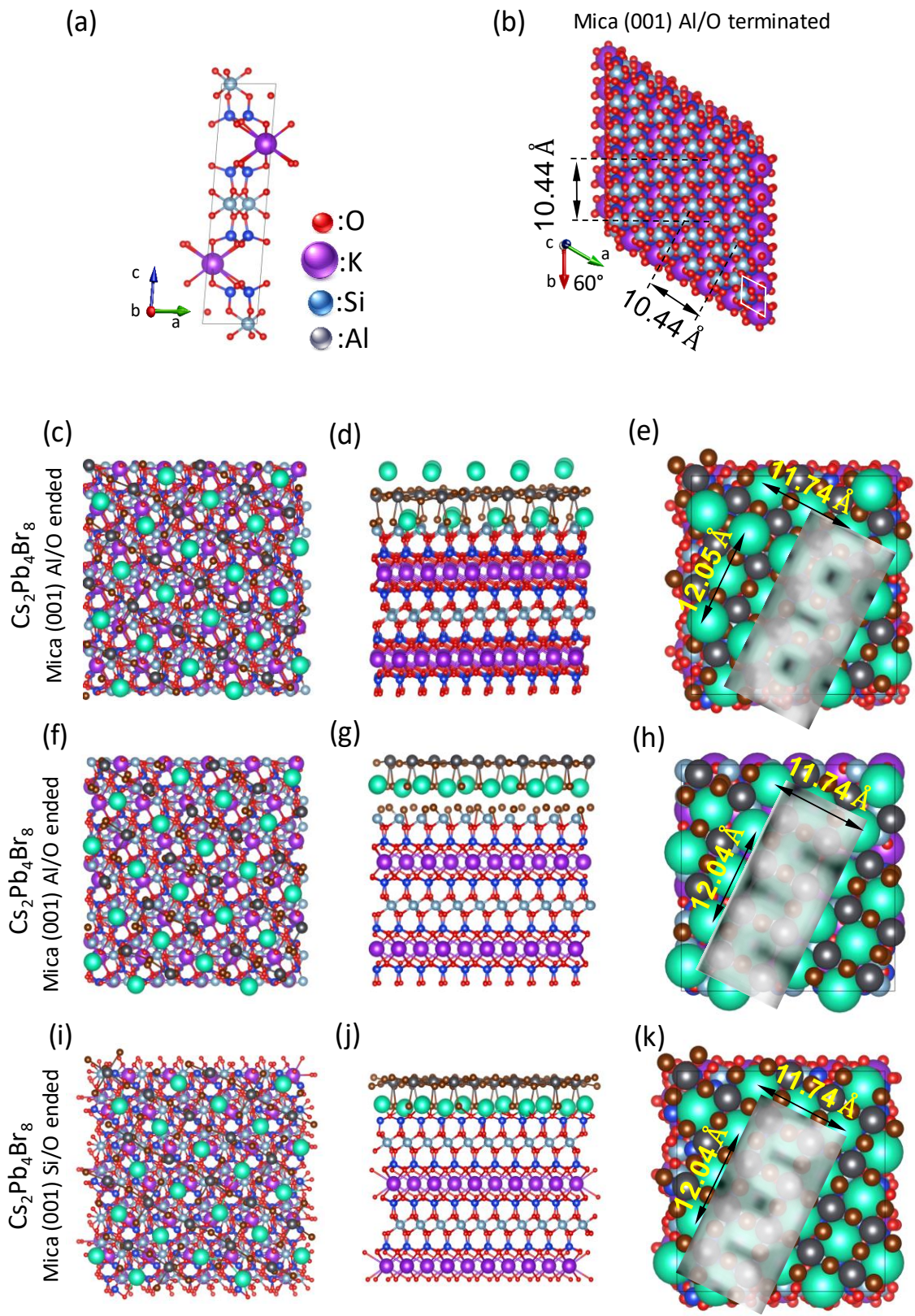

Riedel et al. Figure 6